**RESEARCH**	**Open Access**

# Seasonal variations of EPG Levels in gastro-intestinal parasitic infection in a southeast asian controlled locale: a statistical analysis

Amit K Chattopadhyay[1*] and Subhasish Bandyopadhyay[2]

**Abstract**

We present a data based statistical study on the effects of seasonal variations in the growth rates of the gastro-intestinal (GI) parasitic infection in livestock. The alluded growth rate is estimated through the variation in the number of eggs per gram (EPG) of faeces in animals. In accordance with earlier studies, our analysis too shows that rainfall is the dominant variable in determining EPG infection rates compared to other macro-parameters like temperature and humidity. Our statistical analysis clearly indicates an oscillatory dependence of EPG levels on rainfall fluctuations. Monsoon recorded the highest infection with a comparative increase of at least 2.5 times compared to the next most infected period (summer). A least square fit of the EPG versus rainfall data indicates an approach towards a super diffusive (i. e. root mean square displacement growing faster than the square root of the elapsed time as obtained for simple diffusion) infection growth pattern regime for low rainfall regimes (technically defined as zeroth level dependence) that gets remarkably augmented for large rainfall zones. Our analysis further indicates that for low fluctuations in temperature (true on the bulk data), EPG level saturates beyond a critical value of the rainfall, a threshold that is expected to indicate the onset of the nonlinear regime. The probability density functions (PDFs) of the EPG data show oscillatory behavior in the large rainfall regime (greater than 500 mm), the frequency of oscillation, once again, being determined by the ambient wetness (rainfall, and humidity). Data recorded over three pilot projects spanning three measures of rainfall and humidity bear testimony to the universality of this statistical argument.

## Introduction

Parasitism, or more appropriately gastro-intestinal (GI) parasitism, is arguably the most serious parasitological constraint affecting small ruminants' (e.g. goat, sheep) production worldwide. Economic losses are caused by decreased production, cost of prevention, cost of treatment, and ultimately through the death of infected animals (Barger, 1982; Donald et al., 1982; Perri et al., 2011; Mejla et al. 2011; Cringoli et al., 2004; Mes, 2003; Longo Ribeiro Vilela et al. 2012). The control of GI parasites traditionally relies on grazing management (Perri, et al., 2011), anthelmintic treatment (Costa et al., 2006) or both. However, grazing management schemes are often impractical due to expense or due to the hardiness of infective larvae on pasture. In addition, the evolution of anthelmintic resistance in parasite population threatens the success of drug treatment programmes (Craig, 2006; Prichard, 1990; Waller, 1994; Conder and Campbell 1995; Sangster, 1999). So, the aim of any parasite control program must therefore be to ensure that parasite populations do not exceed threshold levels such that eventual productivity rates are consistently compatible with expected levels of economic production.

GI parasite infection in animals can be quantitatively monitored by the variation in the number of eggs present in faecal samples of livestock over all different seasons of the productive year (Mandal et al., 2010). Forecasting of GI parasitic infection on the basis of meteorological data and computer simulation provides an alternative approach to the control of parasitic infection in a given geographical area (Mackay et al., 1987; Turnbull et al., 1992). The present study is a data based analysis focusing on primary data generated by monitoring the faecal sample

* Correspondence: a.k.chattopadhyay@aston.ac.uk
[1]Aston University, Non-linearity and Complexity Research Group, Engineering and Applied Science, Birmingham B4 7ET, UK
Full list of author information is available at the end of the article examination of animals (cattle and pig) from three Government monitored livestock farms in Meghalaya, India, located in three varying altitudes. The various altitudes are characterized by variations in rainfall and temperature which are influence the parasite build up and thereby affect the variability in observation, recorded through standard deviation estimates of observed data.

The most common gastrointestinal parasite prevalent throughout the year in Meghalaya, India is Strongyle infection. This is because of the high rainfall and humidity prevalent in the North Eastern regions that augments the free living larval development in soil to complete the life cycle of strongyle group of parasite (Bandyopadhyay et al.; 2010, Soulsby, 1982). As the prevalence of free-living larval stages is mainly dependent on micro-climatic environment, therefore, study was initiated to identify the relationship between meteorological parameters and prevalence of gastrointestinal parasitic load in animals. The study would also prove vitally useful in predicting/forecasting the disease occurrence based on the prevailing pattern of rainfall in any specific location.

Three villages at different altitudes in the state of Meghalaya, India were chosen as pilot centres to study the time rate of growth of the infection as a function of the agro-climatic fluctuations. Parallel studies were also conducted to crosscheck the degree of heterogeneity in the spread of the parasitic infection (all three villages show very high rainfall). The collected data clearly indicated much lesser prominence (and importance) of the other two prevalent meteorological parameters (humidity and temperature) in destining the parasite-infected life patterns of the studied animals.

## Materials and methods

A total of 303 cattle and 253 pig faecal samples were collected from the Indian government monitored livestock farms located in Upper Shillong (sub-tropical and temperate zone, altitude 1961 m MSL), Jowai (sub-tropical hill zone, altitude 1380 m MSL) and Kyrdemkulai (mild tropical hill zone, altitude 485 m MSL). All data were collected in complete abeyance of Indian Veterinary Research Institute imposed norms and regulations. These three farms were selected purposefully to cover the high altitude (Upper Shillong), mid altitude (Kyrdemkulai) and low altitude (Jowai) areas.

Meteorological data were collected from the Indian Meteorological department located in these three zones, which were respectively situated at distances of 1 km, 3 km and 5 km from the sample collection sites of Upper Shillong, Jowai and Kyrdemkulai, respectively. Faecal samples were collected per rectal method from each animal of the farm. Samples were divided in to two parts (1 gm each). One part was processed by sedimentation technique (Heath et al., 1968; Carabin et al., 2005; Lloyd et al., 1997) where the eggs with high specific gravity (trematode egg) settled down to the bottom while the lighter ones floated above. The latter part, involving eggs with relatively lower specific gravity, was processed by salt/sugar flotation technique (Smeal et al., 1972; Bello et al., 2009). After processing, the samples were examined under a microscope for identification and quantitative estimation.

The parasitic eggs were identified through characteristic morphological features. For strongyle parasites, the indicators are the ellipsoidal shapes and presence of morula while for the coccidia parasites; the determining features are small round shapes of eggs and four sporocysts. Quantitative estimation was done by counting the number of eggs/ocyst present in the sample (Heath et al., 1968; Carabin et al., 2005; Lloyd et al. 1997 Smeal et al., 1972; Bello et al., 2009).

Based on the collected data, statistical analysis was performed using a combination of mathematical tools based on the programming languages of Matlab, Microsoft Excel and Fortran 90 (all self-written). A major emphasis of this numerical analysis of collected data was to obtain first hand information of the temporal variations of EPG infections pertaining to fluctuations in agro-climatic parameters (rainfall and humidity). Such seasonal fluctuations are depicted in Figures 1, 2, 3, 4, 5, 6, 7, 8, 9, 10, and 11. The figures represent how fast EPG growth (or decay) rate changes with subsequent changes in the rainfall levels. In order to appropriately quantify the individual characteristics of the oscillating pattern, the data (and hence Matlab/Fortran codes) were subdivided in to five different threshold rainfall levels proportional to the amplitude of the oscillation - a) less than 50 mm, b) between 50 mm to 100 mm rainfall, c) between 100 mm to 500 mm, d) between 500 mm to 1000 mm and e) greater than 1000 mm. Each of these threshold demarcations were further subdivided in to smaller windows of varying widths to self-consistently use the real data that was being used. Figures 1, 2, 3, 4, 5, 6, 7, 8, 9, 10 are representative of the changes in the EPG infection pattern with changes in the agro-climatic factors while Figure 11 is a consummate representation of this study that shows the histogram pattern of the EPG infection level with rainfall variations in all three studied villages over all five rainfall threshold levels.

## Results

In this section, we present the aggregate data and a statistical analysis of the same. As previously described, data was collected over a protracted time span of 2 years focusing on three model villages in northeast India that provide the closest conformity to the theoretical prescription presented in this analysis. All three villages are noted for their high rainfall and high humid attributes, the key agro-climatic parameters focused in this study.

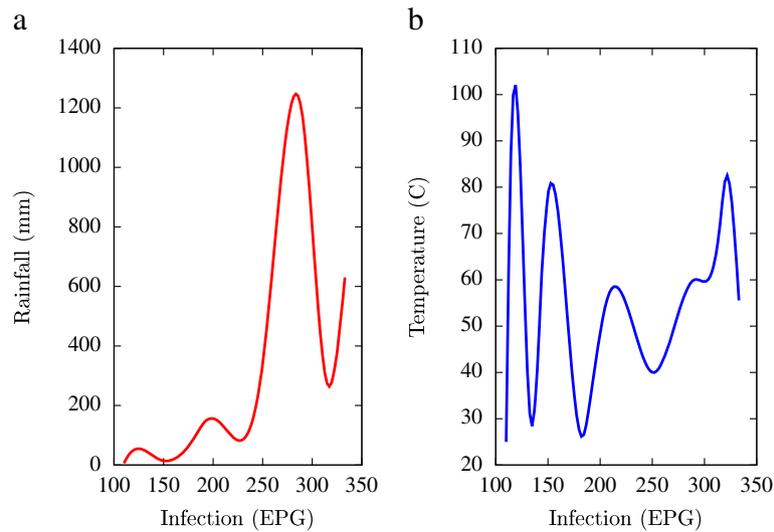

**Figure 1 Fluctuations in the infection level (expressed in units of EPG) due to variations in rainfall (in mm) and temperature (in Centigrade) for Jowai, in the year 2002, are shown respectively in Figure 1a and Figure 1b.** Both rainfall versus infection and temperature versus infection show oscillatory profiles.

As previously indicated, the collected data were analysed using standard statistical methods based on self-written Matlab and Fortran-codes. The study domain was restricted to the linearly stable regime. This is justified since the data itself shows synergy with linear (least square) fits over a broad domain, deviations from which would be the topic of study in our later works. Linear regression fits, including least square fitting methods and time evolution of the resultant histograms form the core of our analysis. In the first of the following two subsections, we focus on the statistical method that was used and the results obtained thereof. These two being largely intertwined, we decided to couple them together. The last subsection summarises the key results obtained and the conclusion that can be drawn from this analysis, with some comments on our ongoing works (focusing on nonlinear mathematical modeling).

## Discussions

Prevalence of gastrointestinal parasitic load in animals measured by egg per gram (EPG) of faeces was found to be linear with the level of precipitation in that region. A vital component of this study was to quantitatively evaluate the comparative importance of the agroclimatic factors in the EPG spread, with particular importance given to rainfall and temperature.

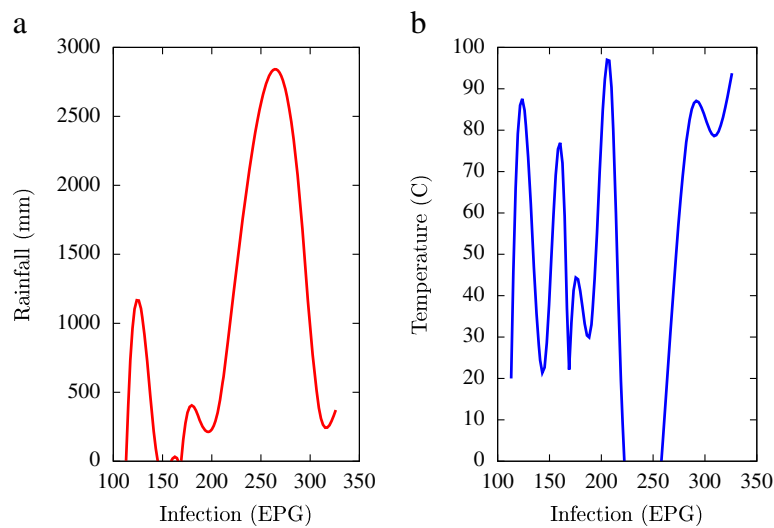

**Figure 2 Fluctuations in the infection level (expressed in units of EPG) due to variations in rainfall (in mm) and temperature (in Centigrade) for Jowai, in the year 2003, are shown respectively in Figure 2a and Figure 2b.** Both rainfall versus infection and temperature versus infection show oscillatory profiles.

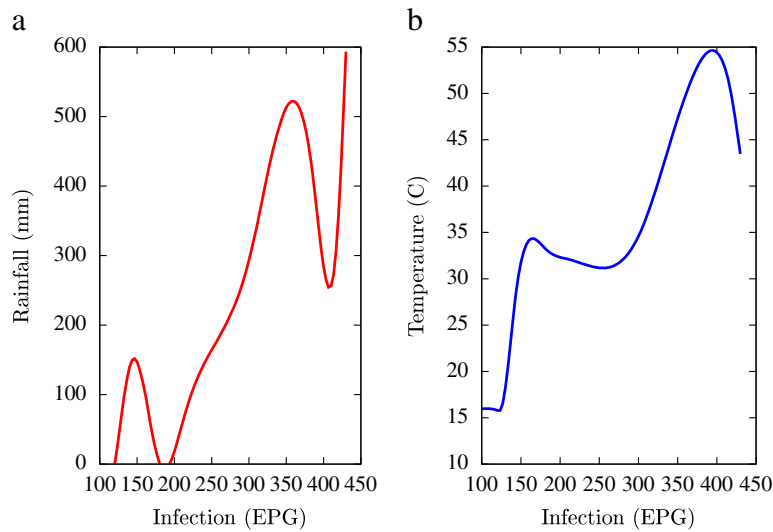

**Figure 3 Fluctuations in the infection level (expressed in units of EPG) due to variations in rainfall (in mm) and temperature (in Centigrade) for Kyrdemkulai, in the year 2002, are shown respectively in Figure 3a and Figure 3b.** Both plots show a jump in the infection level beyond critical values of rainfall and temperature.

All figures have been prepared using the open sourced gnuplot package. Figure 1 shows the fluctuation pattern of the infection level (expressed in units of EPG) due to variations in rainfall (in mm) – Figure 1a - and temperature (in Centigrade) – Figure 1b - for Jowai, in the year 2002. Both plots show oscillatory infection patterns suggesting diurnal dependence. Figure 2 shows similar pattern for the year 2003. Comparative analysis indicates that the specific year of observation does not affect the oscillatory patterns. Figures 3, 4 and 5, 6 show similar statistics for Kyrdemkulai and Upper Shillong respectively. The temperature versus infection plot in Figure 6 indicates that beyond a critical temperature (approximately 30 C), infection grows enormously, although due to seasonal fluctuations in rainfall pattern, an oscillatory (increase–decrease) profile is omnipresent. Figure 7 represents the comparative yearly fluctuations in the infection level (expressed in units of EPG) due to variations in rainfall (in mm) and temperature (in

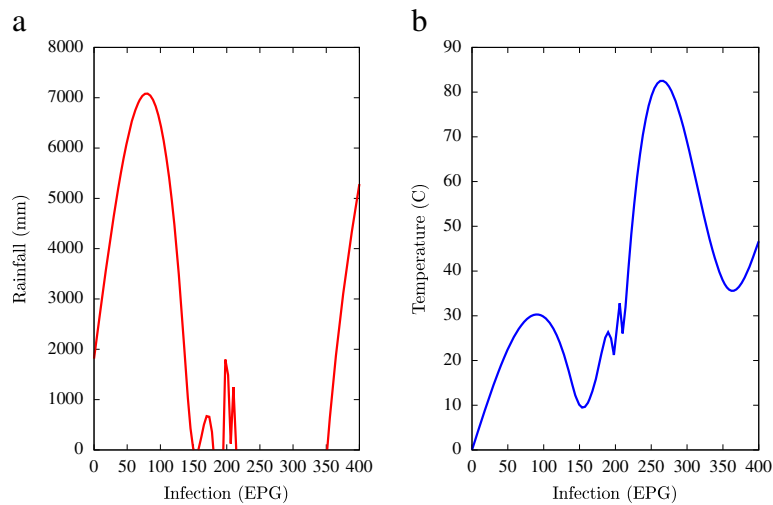

**Figure 4 Fluctuations in the infection level (expressed in units of EPG) due to variations in rainfall (in mm) and temperature (in Centigrade) for Kyrdemkulai, in the year 2003, are shown respectively in Figure 4a and Figure 4b.** Plots indicate oscillatory transitions from a low infection level to a high infection level.

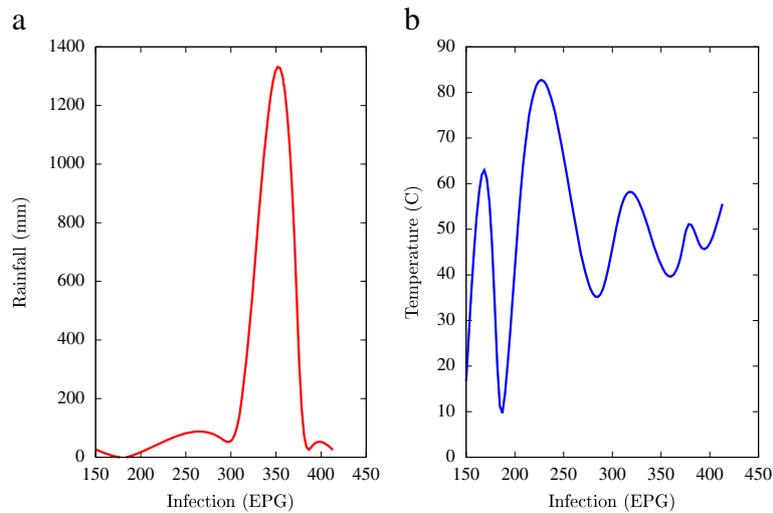

Figure 5 Fluctuations in the infection level (expressed in units of EPG) due to variations in rainfall (in mm) and temperature (in Centigrade) for Upper Shillong, in the year 2002, are shown respectively in Figure 5a and Figure 5b. Both rainfall versus infection and temperature versus infection show decaying oscillatory profiles.

Centigrade) for Jowai. Oscillatory pattern clearly shows a dominating peak in infection level around the monsoon time. Figure 8 portrays comparative yearly fluctuations in the infection level (expressed in units of EPG) due to variations in rainfall (in mm) and temperature (in Centigrade) for Upper Shillong. Oscillatory pattern clearly shows a dominating peak in infection level around the monsoon time, although local highs are always there. Figure 9 represents the fluctuations in the infection level (expressed in units of EPG) due to variations in rainfall (in mm) and temperature (in Centigrade) for Kyrdemkulai. As opposed to Upper Shillong data, shown in Figure 8, the oscillatory pattern in Figure 9 clearly shows a dominating peak in the infection level around the monsoon time, although local highs are always there. Figure 10 is a representative case of comparative estimate of infection (expressed in EPG) variation against changes in the rainfall pattern (expressed in mm) for the three studied zones – Jowai, Upper Shillong and Kyrdemkulai. It is clearly seen that due to heavy rainfall, Kyrdemkulai and Upper Shillong show higher infection levels compared to Jowai. All three zones show oscillatory diurnal infection pattern. Probability density functions of infection level (expressed in EPG) with varying rainfall levels (expressed in mm) give us a picture of the probabilistic change in the infection pattern with rainfall variations (assuming temperature remains unchanged). More specifically, Figure 11 quantifies the probability that infection level will change with variations in the rainfall margin. For this comparison, we have chosen rainfall levels between 0–50 mm, 50–100 mm, 100–500 mm, 500–1000 mm and greater than 1000 mm. The plots clearly indicate the nature of diurnal variation with a sharp peak around monsoon.

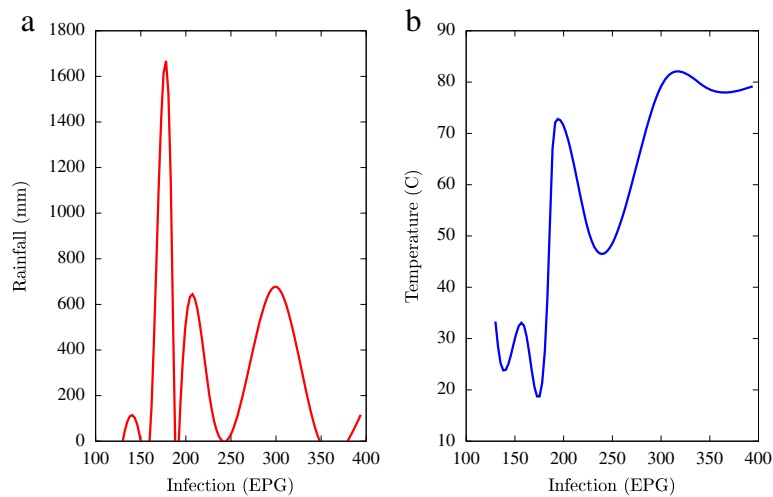

Figure 6 Fluctuations in the infection level (expressed in units of EPG) due to variations in rainfall (in mm) and temperature (in Centigrade) for Upper Shillong, in the year 2003, are shown respectively in Figure 5a and Figure 5b. The temperature versus infection plot (Figure 6b) indicates that beyond a critical temperature (approximately 30 C), infection grows enormously, although due to seasonal fluctuations in rainfall pattern, an oscillatory (increase–decrease) profile is omnipresent.

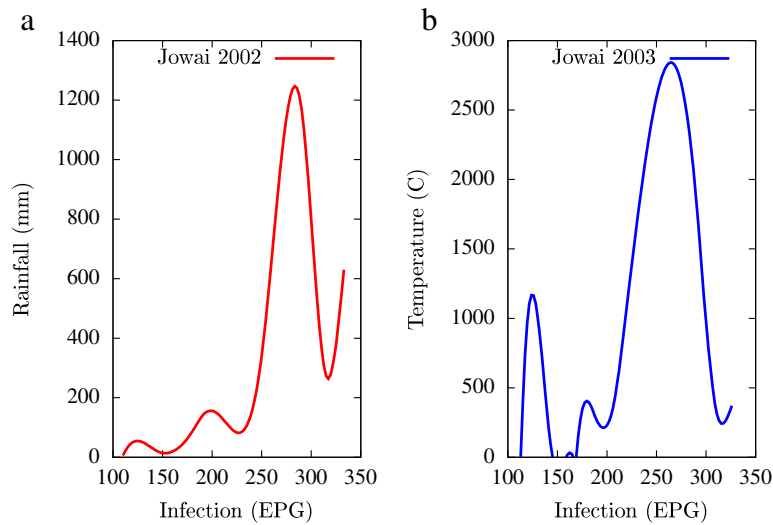

**Figure 7 Comparative yearly Fluctuations in the infection level (expressed in units of EPG) due to variations in rainfall (in mm) and temperature (in Centigrade) for Jowai.** Oscillatory pattern clearly shows a dominating peak in infection level around the monsoon time.

Figure 12 is the crux of our statistical analysis. This is an extrapolated linear fit between infection (expressed in EPG) versus rainfall (expressed in mm) in the $\log_{10}$-$\log_{10}$ scale for 2002 Kyrdemkulai data. The extrapolation clearly suggests a power-law (Pareto) form, where infection grows with rainfall. The precise nature of this growth is given as follows: Infection (EPG) ∼ (Rainfall in mm)$^{0.55}$. It is important to note that although this power law exponent changes slightly if the data (whether from Jowai or Upper Shillong) changes, the change in percentage is never more than 10%. Given this, we are reasonably placed in hypothesizing that the power-law exponent indicates the preponderance of a non sub-diffusive universality class. At present, we are studying possibilities of correlating jumps in the infection levels (Figures 8 & 9) with statistical phase transitions.

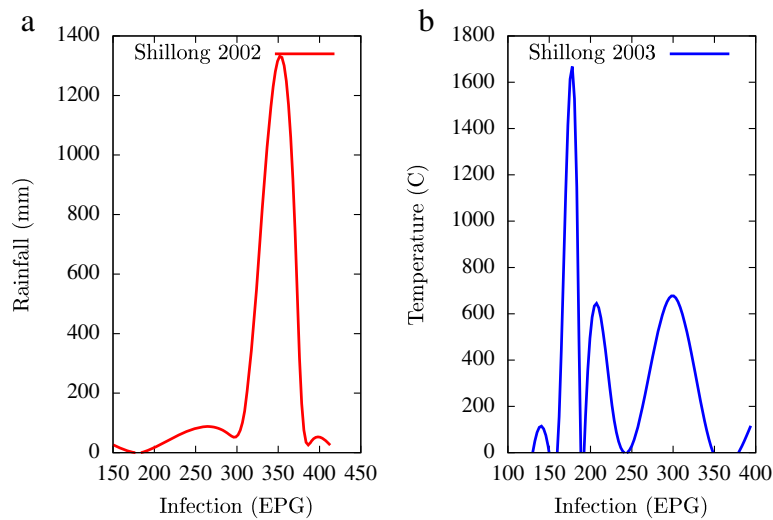

**Figure 8 Comparative yearly fluctuations in the infection level (expressed in units of EPG) due to variations in rainfall (in mm) and temperature (in Centigrade) for Upper Shillong.** Oscillatory pattern clearly shows a dominating peak in infection level around the monsoon time, although local highs are always there.

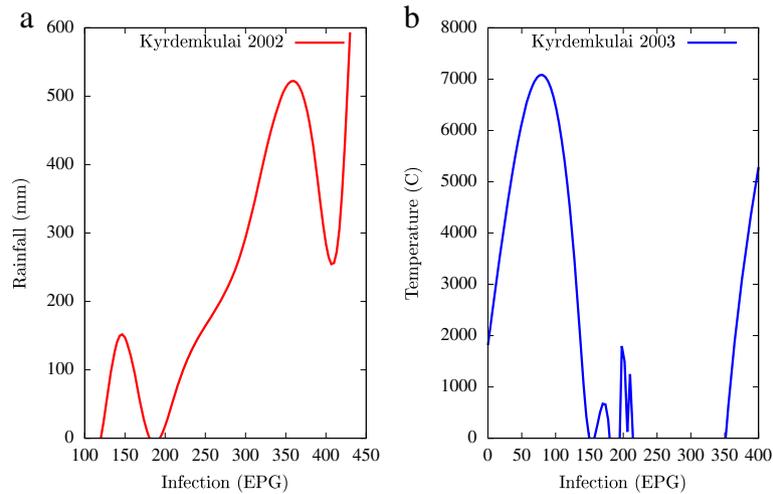

**Figure 9 Comparative yearly fluctuations in the infection level (expressed in units of EPG) due to variations in rainfall (in mm) and temperature (in Centigrade) for Kyrdemkulai.** Oscillatory pattern clearly shows a dominating peak in infection level around the monsoon time, although local highs are always there.

Influence of rainfall on parasitic burden in animals (EPG level in animals) is non-equivocally depicted in all these figures. EPG level was found to be low (50 to 150) when rainfall level was less than 50 mm which was not pathogenic and hence did not directly influence the production level of (infected) animals. Moderate pathogenic effects were found to decrease the production level in animals when rainfall level was 100 to 500 mm (corresponding EPG level 200 to 350). With rainfall levels exceeding 500 mm, severe pathogenic manifestations were recorded in animals with simultaneous increase of EPG over more than 500 (per c.c.) in the faeces of animals.

Irrespective of the year chosen and the region of infection, results (Figures 7, 8, 9) indicate oscillatory seasonal fluctuations for all three regions with peak values for each of these around the monsoon period. The infection spread is seen to be strongly dependent on the amount of rainfall received though. Kyrdemkulai with the highest monsoon rainfall records the lowest possible infection during that period while Upper Shillong with the least rainfall in the aforementioned period shows the strongest effect of infection. There is, however, an underlying component of over simplification in this data based analysis. The effects of temperature on infection, especially during the arid times, can be a vital determining factor that could not be separately studied here (independent of the rainfall effects), a fact that can only be remedied through multivariate analysis of data and/or through model based studies, one of our present works in progress.

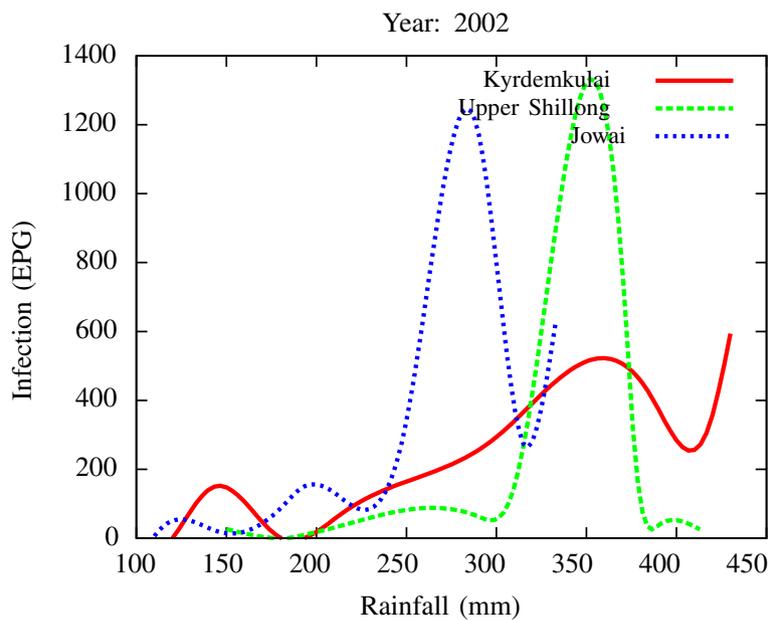

**Figure 10 A comparative estimate of infection (expressed in EPG) variation against changes in the rainfall pattern (expressed in mm) for the three studied zones – Jowai, Upper Shillong and Kyrdemkulai.** It is clearly seen that due to heavy rainfall, Kyrdemkulai and Upper Shillong show higher infection levels compared to Jowai. All three zones show oscillatory diurnal infection pattern.

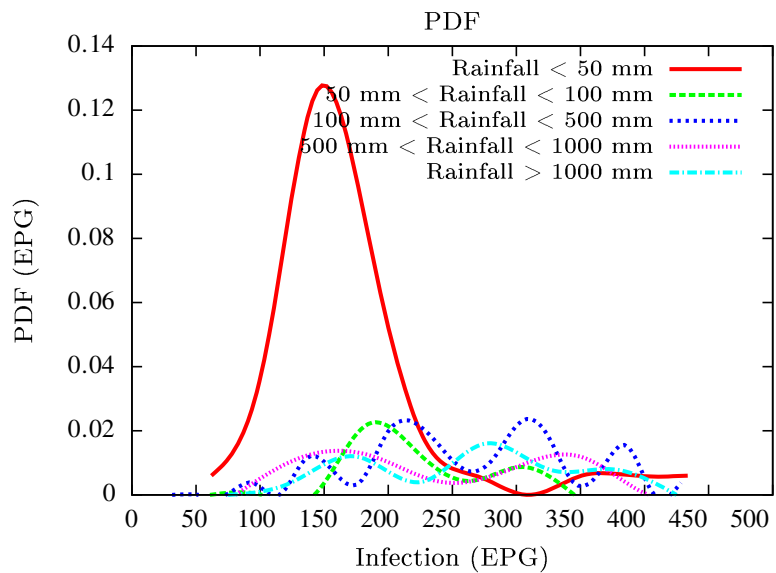

**Figure 11 Probability density functions of infection level (expressed in EPG) with varying rainfall levels (expressed in mm).** We have chosen rainfall levels between 0–50 mm, 50–100 mm, 100–500 mm, 500–1000 mm and greater than 1000 mm. The plots clearly indicate the nature of diurnal variation with a sharp peak around monsoon.

The above analysis gives an impression of the site variation of statistics. In the following, we present the effects of time variation, including seasonal fluctuations. Histograms (splined) obtained from the data were normalized to obtain the respective probability density functions (PDFs) of EPG infection against rainfall in all three regions over the years 2002–2003 across a range of rainfall measures. Clearly low rainfall precipitates a larger infection irrespective of the region of infection (Figure 12). Interestingly, the PDFs too show an oscillatory behavior, clearly indicating the seasonal effects on the infection spread.

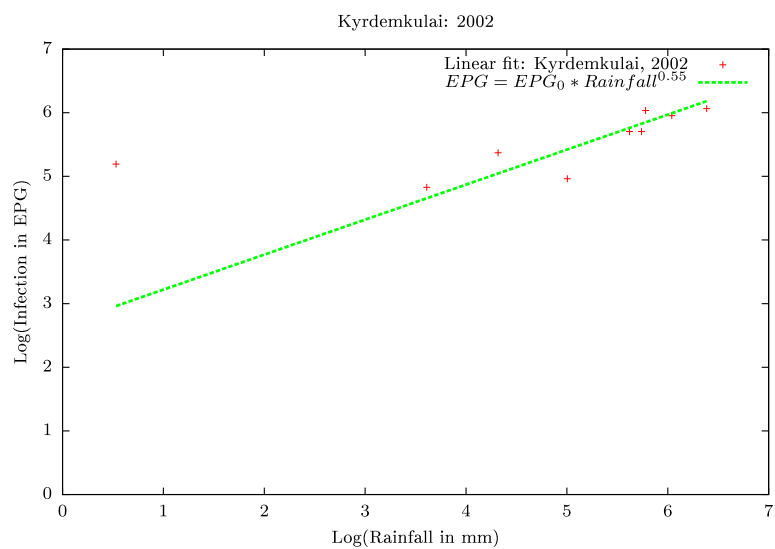

**Figure 12 Linear fit between infection (expressed in EPG) versus rainfall (expressed in mm) in the $\log_{10}$-$\log_{10}$ scale for 2002 Kyrdemkulai data.** The extrapolation clearly suggests a power-law (Pareto) form, where infection grows with rainfall. The precise nature of this growth is given as follows: Infection (EPG) ~ (Rainfall in mm)$^{0.55}$.

To summarise, we conclude that rainfall directly influences the pathogenicity of the parasitic infection level. Unlike other bacterial and viral infections, the gastrointestinal parasites maintain a steady host sustainable relationship, which prevents the host from dying out of this infection. Simultaneously, the parasite contributes to a severe loss of production level in animals through nutrition sharing with the host. Therefore, to prevent production loss, animal parasite controlling anthelmintic treatment should be implemented at the time when precipitation levels of a region are within 100 mm to 500 mm. Our analysis also clearly indicates the presence of seasonal oscillations in the infection level at all points of these cut-off thresholds while also making clear quantitative suggestions contrasting the degree of virility of a strongyle infection against a coccidia infection, with the former type dominating. A lateral derivative of this analysis is the temperature dependence of the infection level (Figures 1, 2, 3, 4, 5, 6) further details of which would be presented through an analysis of a relevant mathematical model in a publication soon to follow.

**Competing interests**
The authors declare that they have no competing interests.

**Authors' contributions**
SB carried out the entire range of livestock related experiments and was responsible for eventual data acquisition in abeyance of regulations imposed by the Indian Veterinary Research Institute. AKC was responsible for the following statistical data analysis and all quantitative formulations quoted in the article. Both authors read and approved the final manuscript.

**Acknowledgment**
SB acknowledges the Director, ICAR Research Complex for the NEH region, Meghalaya for providing necessary infrastructural facilities to carry out parts of this work. SB also acknowledges the Project coordinator, All Indian Network project on Gastrointestinal Parasitism for providing partial financial assistance to conduct this study. AKC acknowledges Royal Society, UK for partial financial support.

**Author details**
[1]Aston University, Non-linearity and Complexity Research Group, Engineering and Applied Science, Birmingham B4 7ET, UK. [2]Eastern Regional Station of Indian Veterinary Research Institute, 37, Belgachia Road, Kolkata 700 037, India.